\documentclass[1p]{elsarticle}
\makeatletter
\def\ps@pprintTitle{%
	\let\@oddhead\@empty
	\let\@evenhead\@empty
	\def\@oddfoot{\centerline{\thepage}}%
	\let\@evenfoot\@oddfoot}
\makeatother

\usepackage{hyperref}

\usepackage{amsmath,amssymb,amsfonts}
\usepackage{algorithmic}
\usepackage{graphicx}
\usepackage{textcomp}
\usepackage{xcolor}
\usepackage{cleveref}
\usepackage{tikz}
\usepackage{amsmath}
\usepackage{url}
\usepackage{subcaption}

\newcommand{\spc}{\,\,\,}

\begin{document}

\begin{frontmatter}

\title{Simulating SQL Injection Vulnerability Exploitation Using Q-Learning Reinforcement Learning Agents}

\author{L{\'a}szl{\'o} Erd{\H o}di}

\cortext[mycorrespondingauthor]{Corresponding author}
\ead{laszloe@ifi.uio.no}

\author{{\AA}vald {\AA}slaugson Sommervoll}
\author{Fabio Massimo Zennaro}

\address{\textit{Department of Informatics} \\
\textit{University of Oslo}\\
Oslo, Norway}

\begin{abstract}
In this paper, we propose a formalization of the process of exploitation of SQL injection vulnerabilities. We consider a simplification of the dynamics of SQL injection attacks by casting this problem as a security capture-the-flag challenge. We model it as a Markov decision process, and we implement it as a reinforcement learning problem.
We then deploy reinforcement learning agents tasked with learning an effective policy to perform SQL injection; we design our training in such a way that the agent learns not just a specific strategy to solve an individual challenge but a more generic policy that may be applied to perform SQL injection attacks against any system instantiated randomly by our problem generator.
We analyze the results in terms of the quality of the learned policy and in terms of convergence time as a function of the complexity of the challenge and the learning agent's complexity.
Our work fits in the wider research on the development of intelligent agents for autonomous penetration testing and white-hat hacking, and our results aim to contribute to understanding the potential and the limits of reinforcement learning in a security environment.
\end{abstract}

\begin{keyword}
SQL injection\sep capture the flag\sep vulnerability detection\sep autonomous agents\sep reinforcement learning\sep Q-learning
\end{keyword}

\end{frontmatter}


\section{Introduction}
SQL injection is one of the most severe vulnerabilities on the web. It allows attackers to modify the communication between a web server and a SQL database by sending crafted input data to the website. By controlling the SQL query instantiated by a server-side script, attackers can extract from a database data that they should not normally be authorized to retrieve. In extreme cases, attackers can persistently change the databases or even exploit the SQL injection vulnerability to send remote commands for execution on the website server. To secure a system, detecting SQL injection vulnerabilities is a crucial task for ethical hackers and legitimate penetration testers.

In this paper, we consider automatizing the process of exploiting SQL injection vulnerability through machine learning. In particular, we assume that a vulnerability has been identified, and then we rely on reinforcement learning algorithms to learn how to exploit it. Reinforcement learning algorithms have been proved to be an effective method to train autonomous agents to solve problems in a complex environment, such as games \cite{mnih2015human,silver2017mastering,vinyals2019grandmaster}. Following this methodology, we cast the problem of exploiting SQL injection vulnerabilities as an interactive game. In this game, an autonomous agent probes a system by sending queries, analyzing the answer, and finally working out the actual SQL injection exploitation, much like a human attacker.
In this process, we adapt the problem of exploiting SQL injection vulnerabilities to the more generic security-related paradigm of a capture-the-flag (CTF) challenge. A CTF challenge constitutes a security game simulation in which ethical hackers are required to discover a vulnerability on a dedicated system; upon discovering a vulnerability, an ethical hacker is rewarded with a flag, that is, a token string proving her success. It has been proposed that the generic CTF setup may be profitably used to model several security challenges at various levels of abstraction \cite{erdodi2020agent}. We implicitly rely on this framework to model SQL injection exploitation as a CTF problem and map it to a reinforcement learning problem.
Concretely, we implement a simplified synthetic scenario for SQL injection exploitation, we deploy two standard reinforcement agents, and we evaluate their performance in solving this problem. Our results will provide a proof of concept for the feasibility of modeling and solving the exploitation of SQL injection vulnerability using reinforcement learning.

This work fits in the more general line of research focused on developing and applying machine learning algorithms to security problems. Reinforcement learning algorithms have been previously considered to tackle and solve similar penetration testing problems, although they have never been applied specifically to the problem of SQL injection vulnerability exploitation. In particular, generic penetration testing has been modelled as a reinforcement problem in \cite{sarraute2013penetration,bland2020machine, ghanem2020reinforcement,pozdniakov2020smart}, while explicit capture-the-flag challenges have been considered in \cite{zennaro2020modeling}. Moreover, autonomous agents were invited to compete in a simplified CTF-like challenge in the DARPA Cyber Grand Challenge Event hosted in Las Vegas in 2016 \cite{CyberGrand}.

\section{Background}
In this section, we review the main ideas from the field of security and machine learning relevant to the present work.

\subsection{SQL Injection} \label{sec:SQLinjection}
Dynamic websites are widespread nowadays. To provide a rich user experience, they have to handle a large amount of data stored for various purposes, such as user authentication. Access to the stored data, as well as their modification, has to be very fast, and an effective solution is to rely on relational databases such as \textit{mysql}, \textit{mssql}, or \textit{posgresql}. All these database systems are based on the standard query language \emph{SQL} (Structured Query Language).

SQL communication between the website and the SQL server consists of SQL queries sent out by the web server and SQL responses returned by the SQL server. The most frequently used operation is data retrieval using the $\mathtt{SELECT}$ command along with a $\mathtt{WHERE}$ clause to select columns and rows from a table satisfying a chosen condition; in advanced statements, multiple SQL queries can be concatenated with a $\mathtt{UNION}$ statement returning one table made up by the composition of the query answers. A simple example of a query where two columns are selected from a table with filtering using the value of the third column is:
\[
\mathtt{SELECT \spc Column1,Column2 \spc FROM \spc Table1 \spc WHERE \spc Column3 = 4}.
\]
A more complex example where two query results are concatenated with the $\mathtt{UNION}$ keyword is:
\begin{align*}
& \mathtt{SELECT \spc Column1 \spc FROM \spc Table1 \spc WHERE \spc Column2 = 4 \spc UNION} \\
& \mathtt{SELECT \spc Column4 \spc FROM \spc Table2 \spc WHERE \spc Column5 >= 12}.
\end{align*}


SQL injection happens when the server side script has an improperly validated input that is inserted into the SQL query directly or indirectly by the server side script. Because of the inappropriate validation, the attacker can gain a full or partial control over the query. In the easiest case the attacker can modify the expression evaluation in the $\mathtt{WHERE}$ clause of the query by escaping from the input variable and adding extra commands to the query. For example, if the SQL query exposed by the script is:
\[
\mathtt{SELECT \spc Column1 \spc FROM \spc Table1 \spc WHERE \spc Column2 = \spc} \textnormal{\textit{input1}},
\]
then, using the misleading input $\mathtt{1 \spc OR \spc 1=1}$, the $\mathtt{WHERE}$ clause evaluation will be always true independently of the first condition:
\[
\mathtt{SELECT \spc Column1 \spc FROM \spc Table1 \spc WHERE \spc Column2 = 1 \spc OR \spc 1=1}.
\]
Note that in the previous example, the SQL engine behaves as if the input data was $\mathtt{1}$, and $\mathtt{OR \spc 1=1}$ was part of the pre-existing query. In more refined injections, the attacker can add a $\mathtt{UNION}$ statement and craft two queries from the original single query. In these cases, the attacker must find or guess the number and type of the selected columns in the second query to align with the first query.

Overall, the process of exploitation of an SQL injection vulnerability can be decomposed into the following non-ordered steps based on conventional exploitation logic:

\begin{enumerate}
\item \emph{Finding a vulnerable input parameter}: a website can accept multiple parameters with different methods and different session variables. The attacker has to find an input parameter that is inserted in a SQL query by the script with missing or improper input validation.

\item \emph{Detecting the type of the vulnerable input parameter}: the attacker has to escape from the original query input field. For instance, if the input parameter is placed between quotes by the script, then the attacker has to also use a quote to escape from it; if the original query were, for example, $\mathtt{SELECT \spc Column1 \spc FROM \spc Table1} \allowbreak \mathtt{WHERE \spc Column2 = \spc} \mathtt{'}\textnormal{\textit{input1}}\mathtt{'}$, then the escape input has to also start with a quote: $\mathtt{1' \spc OR \spc '1'='1}$. Note that here, in the added Boolean comparison, two strings $\mathtt{'1'}$ are compared, but the closing quote of the second string is missing because it is placed there by the original script itself.

\item \emph{Continuing the SQL query without syntax errors}: escaping from the input provides options for the attacker to continue the query. The SQL syntax has to be respected, considering possible constraints; for instance, escaping from the string requires inserting a new string opening quote at the end of the input. A common trick is to use a comment sign at the end of the input to invalidate the rest of the SQL query in the script.

\item \emph{Obtaining the SQL answer presentation in the HTTP response}: after submitting her SQL query, the attacker obtains an answer through the website. Despite the SQL engine answering with a table, this raw output is highly unlikely to be visible. The generated HTTP answer with the HTML body delivered to the attacker is a function of the unknown SQL query processing by the server-side code. In some cases, the attacker can see one or more fields from the SQL answer, but in other cases, the query result is presented only in a derived form by different HTML responses. In this latter case, the attacker can carry out a Boolean-based blind SQL injection exploitation by playing a \textit{true} or \textit{false} game with the website. 

\item \emph {Obtaining database characteristics for advanced queries}: To insert meaningful queries in the original input, the attacker needs to uncover the names of tables or columns. This can require to select values from the \textit{information schema} table in advance. If the attacker aims to use the $\mathtt{UNION \spc SELECT}$ approach, she has to obtain the column count and types of the first query in order to be aligned with the first query.

\item \emph{Obtaining the sensitive information}: once she knows the necessary parts of the original query (input type, structure of the query) and having all information about the databases (database names, table names, column names), then the attacker can obtain the required confidential data.

\item \emph{Carrying out extra operations}: in addition to retrieving data from the database, the attacker can carry out extra tasks such as writing a script file to the server using the $\mathtt{SELECT \spc INTO \spc outfile}$ command. This type of advanced exploitation is above the normal aim of SQL injection exploitation, and the objective is often to create a remote command channel for the attacker for further attacks.
\end{enumerate}

Notice that these steps are not necessarily taken out in this order: an attacker may skip or repeat steps, according to the attack she is mounting.


\subsection{Reinforcement Learning}

Reinforcement learning \cite{sutton2018reinforcement} constitutes a family of machine learning algorithms designed to solve problems modeled as \emph{Markov decision processes} (MDP).

A MDP allows to describe the interaction of an \emph{agent} with an \emph{environment} (or \emph{system}). The aim of the agent is to learn an effective \emph{policy} $\pi$ by probing and interacting with the system.
Formally, the environment is defined as a tuple:
\[
\left\langle \mathcal{S},\mathcal{A},\mathcal{T},\mathcal{R}\right\rangle
\]
where $\mathcal{S}$ is a set of states in which the system can be, $\mathcal{A}$ is a set of actions the agent can take on the system, $\mathcal{T}: \mathcal{S} \times \mathcal{A} \rightarrow \mathcal{S}$ is a (deterministic or probabilistic) transition function defining how the system evolves from one state to the next upon an action taken by the agent, and $\mathcal{R}: \mathcal{S} \times \mathcal{A} \rightarrow \mathbb{R}$ is a (deterministic or probabilistic) reward function returning a real-valued scalar to the agent after taking a certain action in a given state. The model is \emph{Markovian} as its dynamics in state $h_i\in\mathcal{S}$ depends only on the current state and not on the history; alternatively, $h_i\in\mathcal{S}$ constitutes a sufficient statistic of the history of the system to determine its dynamics.

Reinforcement learning in this MDP setting is formally defined as the learning of an optimal action policy $\pi^*(a\vert h)=P(a\vert h)$ that determines a distribution of probability over actions $a$ in a given state $h$, and such that the long-time expected sum of rewards over a time horizon $T$ is maximized:
\[
\pi^{*}=\arg\max_\pi G_t = \arg\max_\pi \sum_{t=0}^{T}\gamma^{t}E_{\pi}\left[r_{t}\right],
\]
where $\gamma$ is a discount factor (introduced for mathematical and modelling reasons), $E_p[\cdot]$ is the expected value with respect to distribution $p$, and $r_t$ is the reward obtained at time-step $t$.

A reinforcement learning agent learns (to approximate) an optimal policy $\pi^*$ relying on minimal prior knowledge encoded in its algorithm. The simplest model-free agents are provided simply with the knowledge of the feasible action set, and they learn their policy by interacting with the environment, observing their rewards, and estimating the value of different actions in different states. Undertaking an action and observing its result is called a \emph{step}; a collection of steps from the initial state of the MDP to a final state of the MDP (if it exists) or to an arbitrary termination condition (e.g., the maximum number of steps) is called an \emph{episode}. Several algorithms are presented in the reinforcement learning literature \cite{sutton2018reinforcement}; we will briefly review the algorithms relevant to this paper in the next section.

\subsection{Literature overview}
Machine learning has recently found application in many fields in order to solve problems via induction and inference, including security \cite{StasinopoulosAutomatic}. Success in complex tasks like image recognition \cite{krizhevsky2012imagenet} or natural language processing \cite{vaswani2017attention} has spurred the application of supervised deep neural networks to security-related problems where data is abundant; examples include processing code to detect vulnerabilities \cite{russell2018automated} or malware \cite{tobiyama2016malware}. However, the supervised paradigm fits less well a dynamic problem such as web vulnerability exploitation, where multiple decisions and actions may be required to achieve the desired goal. In this context, a more suitable solution is offered by reinforcement learning algorithms designed to train an agent in a complex environment via trial-and-error. Remarkable successes on games like Go \cite{silver2017mastering} or Starcraft II \cite{vinyals2019grandmaster} suggest that this approach may be fruitfully applied to web vulnerability exploitation or penetration testing in general. Applications of machine learning and reinforcement learning algorithms for offensive security has seen application in the 2016 Cyber Grand Challenge 2016 \cite{CyberGrand}, a competition in which participants were requested to deploy automatic agents to target generic system vulnerabilities.

To the best of our knowledge, machine learning has been used so far only to detect SQL injection vulnerabilities, but never for exploitation. Several papers have indeed considered the problem of detecting SQL injection using machine learning methods.
In \cite{skaruz2007recurrent} recurrent neural networks were trained to detect and discriminate offensive SQL queries from the legitimate ones. \cite{joshi2014sql} proposed a classifier that uses a combination of Naïve Bayes modules and Role Based Access Control mechanisms for the detection of SQL injection. \cite{singh2015Sql} implemented an unsupervised clustering algorithm to detect SQL injection attacks.
\cite{ogbomon2017applied} showed a proof-of-concept implementation of a supervised learning algorithm and its deployment as a web service able to predict and prevent SQL injection accurately.
\cite{ross2018sql} exploited network device and database server logs to train neural networks for SQL injection detection. \cite{hasan2019detection} tested and compared 23 different machine learning classifiers and proposed a model based on a heuristic algorithm in order to prevent SQL injection attacks with high accuracy.
\cite{tang2020detection} also presented a SQL injection detection method based on a neural network processing simple eight-features representations and achieving high accuracy.
All these works demonstrate the interest of the community in the problem of dealing with SQL injection vulnerabilities.
However, most of these studies have focused on the problem of identifying the vulnerability, and they have relied on supervised or unsupervised machine learning algorithms. The most complex part of a SQL injection attack, the exploitation, has not been considered for automation. Our work aims at filling this gap, providing a formalization and an implementation of a reinforcement learning model targeted at the problem of exploiting a SQL injection vulnerability.

\section{Model}
This section describes how we modeled the problem of performing SQL injection as a \emph{game} that can be tackled with standard RL methods.

\subsection{Simplification of the SQL problem}
Based on the number of possibilities, solving a general SQL injection problem with RL would require considering numerous types of actions and a high number of states. Although the final aim is to solve such an arbitrary SQL injection exploitation problem, here our approach is to consider a scenario with the following simplifications.

\emph{Capture the Flag Jeopardy-style problems} - In case of real attacks involving SQL injection exploitation, the attacker might be driven by vague objectives (e.g., eliciting information, writing a script to the server).
Moreover, the attacker should consider the presence of a possible defense team; her attacks might be observed, and counteractions might be taken; in such a case, covering her tracks might increase the chances of success.
In our solutions, we model the problem as a \emph{Jeopardy-style Capture the Flag} (CTF) game, in which the environment is static (no defensive blue team), and the victory condition is clearly defined in the form of a flag (no fake flags are provided, and the agent can easily identify the flag).

\emph{Only one vulnerable input for the website} - Finding the vulnerable input for the SQL injection exploitation can also be challenging. In an average website, numerous input data can be sent using different HTTP methods. Access right for the pages complicates the case as well.
Our focus is on the exploitation of the vulnerability, not on the vulnerable parameter finding. As such, we consider a mock website that has only one input parameter that is vulnerable to SQL injection. Note that this does not mean that any other characteristic of the vulnerability is known to the agent. The idea is to avoid repeatedly sending the same input for all input parameters of the website.

\emph{No input validation by the server side script} - When the client sends the input parameter, the server side script can modify or customize it. This processing may completely prevent the possibility of a SQL injection or limit the options of the attacker. In our simplified approach, we assume that the input data is placed directly into the SQL query without any transformation. We also assume that there is only one query in the server side script, and the input does not go through a chain of queries.

\emph{The SQL result can be presented in different ways} - Similar to the input transformation, the server side script is responsible for processing the output SQL answer and generating the web HTML answer. Different degrees of processing are possible, ranging from minimal processing (showing the actual response embedded in the HTML page) to a complete transformation (returning a different web page according to the result of the query without embedding the actual data in the page).
We consider a representation that strikes a balance between simplicity and realism, in which the answer web page contains fields from the queried table, thus providing the agent with an indication of success or failure.

\emph{Unified table and column names} - During the exploitation, the attacker has to identify different databases with different table names, and it might be necessary to map the table characteristics such as column names and types. We consider only one background database with unified names for tables and columns.

\emph{Only three data types in the tables} - We consider three different data types: integer, varchar (string), and datetime to simplify the complexity of the problem.

\emph{Union is allowed with column matching} - Our exploitation strategy can use the $\mathtt{UNION}$ statement to concatenate query results. We assume that this is allowed by the SQL server with the only condition to have the same number of results in the columns in both queries.

\emph{No error messages} - In some cases, the SQL errors are presented in the web answer. Using the table names or column names leaked by the SQL error can help the attacker. In our assumption, we consider that the SQL error messages are not visible to the attacker.

Because of the above simplifying assumptions, our agent will not consider all the possible types of actions that could take place during an SQL injection exploitation. Indeed, with respect to the different SQL injection steps that we have identified in Section \ref{sec:SQLinjection}, our agent will focus on the problems of detecting the type of the input parameter (step 2), formulating a syntactically correct query (step 3), and obtaining database characteristics for advanced $\mathtt{UNION \spc SELECT}$ queries (step 5) in order to obtain the sensitive information (step 6). We assume that the identity of the vulnerable parameter is known (step 1) and that the presentation of the SQL answer is transparent (step 4). We do not consider the problem of carrying out extra operations (step 7).

\subsection{Reinforcement learning modelling}
In order to deploy reinforcement learning agents to perform SQL injection exploitation, we model our problem as an MDP. We take the potential attacker or pentester as the reinforcement learning agent, and we represent the vulnerable webpage with its associated database as the MDP environment.

\paragraph*{MDP}
We map the set of state $\mathcal{S}$ to the states of the webserver. Since we modeled the problem as a static CTF challenge, we assume a system whose underlying behavior does not change upon the sending of requests by the agent (e.g., there are no mechanisms in place to detect a possible attack and modify the dynamics of the server); formally our webserver is stateless, meaning that it has just a singleton state. However, in order to track the knowledge of the agent (which actions have been attempted and which results were produced), we account in the state variable also for the knowledge accumulated by the agent. Therefore, our states are defined by the history $h_i$ of actions taken (and responses seen) by the agent. Clearly, such a state guarantee that our system is Markovian as it trivially summarizes the entire history of the interactions between the agent and the system.
We map the set of actions $\mathcal{A}$ to a (finite) set of SQL string that the agent can send to the webpage.
We map the transition function $\mathcal{T}$ to the internal logic that drives the webpage. Since the system is stateless in our simulation, this function is simply an identity mapping over the singleton.
Finally, we map the reward function $\mathcal{R}$ to a signal that returns a positive feedback when the agent performs the SQL injection and retrieves the flag and a negative feedback for each unsuccessful query it sends.

\paragraph*{RL agents}
In order to actually solve the MDP defined above, we consider two different concrete algorithms for our agent.

The first algorithm is the standard \emph{tabular Q-learning} \cite{sutton2018reinforcement}. Q-learning is a value-based algorithm that aims at deriving an optimal policy $\pi^*$ by estimating the value of each action in any possible state:
\[
Q\left(a_{j},h_{i}\right)=E_{\pi}\left[G_{t}\vert a_{t=j},h_{t=i}\right],
\]
that is, the Q-value of action $a_j \in \mathcal{A}$ in state $h_i \in \mathcal{S}$ is the long-term expected reward $G_t$ under the current policy $\pi$ assuming that at the current step $t$ we were in state $h_i$ and had undertaken action $a_j$.
The estimated Q-values are updated at run-time, step after step, using a temporal-difference algorithm, that is, progressively correcting the difference between the current estimates of the agent and the actual reward it obtains:
\[
Q\left(a_{t},h_{t}\right)\leftarrow Q\left(a_{t},h_{t}\right)+\eta\left[r_{t}+\gamma\max_{a}Q\left(a,h_{t+1}\right)-Q\left(a_{t},h_{t}\right)\right],
\]
where $\eta\in\mathbb{R}_{>0}$ is a learning rate.
An action policy may be simply defined by choosing, in state $h_t$, the action $a^*$ that guarantees the highest Q-value $Q(a^*,h_t)$. However, at learning time, such a greedy policy may lead the agent not to explore all its possibilities exhaustively; therefore, it is common to introduce an exploration parameter $\epsilon\in[0,1]$, and define the agent policy as:
\[
a_{t}=\begin{cases}
a^{*}=\arg\max_{a}Q\left(a,h_{t}\right) & \textrm{with probability }(1-\epsilon)\\
\sim\mathsf{Unif}(\mathcal{A}) & \textrm{otherwise },
\end{cases}
\]
that is, we choose the optimal action $a^*$ with probability $1-\epsilon$, otherwise we sample uniformly at random an action from the action set $\mathcal{A}$.
In the tabular Q-learning algorithm, the estimated Q-values are explicitly encoded in a table. This algorithm is guaranteed to converge to the optimal solution; however, such a representation may not scale well with the dimensionality of the action and state space.

The second algorithm we consider is the \emph{deep Q-learning} \cite{mnih2015human}. A deep Q-learning (DQN) agent aims at estimating Q-values like the first agent, but instead of instantiating a matrix, it relies on a (deep) neural network to approximate the Q-values. The use of a neural network avoids the allocation of large matrices, thus allowing to deal with large action and state spaces; however, the ensuing approximation makes it more challenging to interpret the learned model and to guarantee convergence \cite{sutton2018reinforcement}.

\section{Experimental simulations}
In this section, we describe the environment we developed, and then we present our simulations and their results. All our simulations are publicly available online at \url{https://github.com/FMZennaro/CTF-SQL}.

\subsection{Environment}
Our environment consists of a simplified scenario in which an agent interacts with a web page by sending a parameter in the form of a string $s$; the web page embeds the parameter in a pre-generated SQL query and returns the result of the execution of such a query to the agent.

\paragraph*{Database}
In our problem, we assume that the web page interacts with a randomly generated database composed of $N_t>1$ tables; for simplicity, all tables are named $\mathtt{Table}i$, where $i$ an index between $1$ and $N_t$ (e.g.: $\mathtt{Table2}$). Each table is defined by a random number $N_c>1$ of columns with a random data type chosen among \emph{integer}, \emph{string} or \emph{datetime}; all columns are given default names with the form $\mathtt{Column}j$, where $j$ an index between $1$ and $N_c$ (e.g.: $\mathtt{Column3}$). Each table is populated by a random number of rows $N_r>1$ containing data fitting the data type of each column.
Additionally, we instantiate one final table named $\mathtt{Flagtable}$ with a single \emph{string} column named $\mathtt{flag}$, and containing only one record with the string "flag".

In our simulation, in every episode we sample uniformly at random $N_t,N_c,N_r$ in the interval $[1,5]$. Notice, however, that the complexity of the problem as it is defined below depends only on $N_c$.

\paragraph*{Pre-generated SQL query}
The pre-generated SQL query on the web server that accesses the database is instantiated randomly at the beginning of each episode, and it can take the following general form:
\[
\mathtt{SELECT} \spc [Columns] \spc \mathtt{FROM} \spc [Table] \spc \mathtt{WHERE} \spc [Column][Condition][Input],
\]
where:
\begin{itemize}
   \item $[Columns]$ is a list of $n>1$ columns;
    \item $[Table]$ is the name of a table;
    \item $[Column]$ is the name of a column;
    \item $[Condition]$ is a logical operator chosen in the set $\{ =, >, \mathtt{BETWEEN \spc '01/01/2000 \spc 12:00:00 \spc AM' \spc AND}\}$;
    \item $[Input]$ is the user defined string $s$ which may take one of the following forms $\{ s, "s", 's'\}$.
\end{itemize}
A pre-generated SQL query would be, for instance:
\[
\mathtt{SELECT \spc Column3,Column4 \spc FROM \spc Table2 \spc WHERE \spc Column1 = 's'}
\]

\paragraph*{SQL injection}
The learning agent is not aware of the specific pre-generated SQL query running on the web page, and it can only discover the possible vulnerability by sending strings $s$ and observing the result.

However, we assume that the agent is aware of the generic form of the SQL query, which means that it knows that the SQL injection solution would have the generic form:
\[
[Escape] \spc \mathtt{UNION \spc SELECT} \spc [FColumns] \spc \mathtt{FROM \spc Flagtable\#},
\]
where:
\begin{itemize}
    \item $[Escape]$ is an escape character introducing the SQL injection, and which must be chosen in the set $\{ \epsilon, ", '\}$, where $\epsilon$ is the empty string;
    \item $[FColumns]$ is the repetition of a dummy column name (e.g.: $\mathtt{flag}$) ranging over the number of columns in the pre-generated SQL query.
\end{itemize}
Notice that the hash symbol $\#$ at the end is introduced for generality to comment out any other following instruction.
As an illustration, the SQL injection query for the example pre-generated SQL query above would be:
\[
\mathtt{' \spc UNION \spc SELECT \spc flag,flag \spc FROM \spc Flagtable}
\]

\paragraph*{Action space}
The knowledge of the generic form of the solution allows us to identify two sub-objectives for the agent: (i) identify the correct escape character, and (ii) guess the correct number of columns to insert in the SQL injection string. With reference to the SQL injection steps in Section \ref{sec:SQLinjection}, notice that task (i) is related to step 2 and 3 of SQL exploitation, while task (ii) is related to step 5 of SQL exploitation.

Based on the identification of these sub-objectives, we can define a finite and restricted set of actions that would allow the agent to achieve its sub-goals and send to the web server the right query $s$ to perform the exploit.
More specifically, the action set $\mathcal{A}$ can be partitioned into three sub-sets of conceptually different actions.
The first subset contains \emph{escape actions} $\mathcal{A}_{esc}$, that is, queries aimed at simply discovering the right escape characters needed for the SQL injection.
This set contains the following actions:
\begin{align*}
\mathcal{A}_{esc} = & \{ \mathtt{" \spc and \spc 1=1\#}, \\
& \mathtt{" \spc and \spc 1=2\#},\\
& \mathtt{' \spc and \spc 1=1\#},\\
& \mathtt{' \spc and \spc 1=2\#},\\
& \mathtt{and \spc 1=1\#},\\
& \mathtt{and \spc 1=2\#} \}.
\end{align*}
The cardinality of this subset is $\left| \mathcal{A}_{esc} \right| = 3 \cdot 2 = 6$, that is two actions for each one of the three possible escape solutions.

The second subset contains \emph{column actions} $\mathcal{A}_{col}$, that is, queries that can be used to probe the number of columns necessary to align the $\mathtt{UNION \spc SELECT}$ to the original query. This set of actions contains queries with the generic form:
\[
\mathcal{A}_{col} = \{
[Escape] \spc \mathtt{UNION \spc SELECT} \spc [Columns] \spc [Options]\# \},
\]
where $[Columns]$ corresponds to a list of a variable number, between $1$ and $N_c$, of columns, and $[Options]$ are output formatting options using the SQL commands $\mathtt{LIMIT}$ and $\mathtt{OFFSET}$. The options commands do not affect the results of the query in this current simulation.
The cardinality of this subset is $\left| \mathcal{A}_{col} \right| = 3 \cdot 10 = 30$, that is ten actions for each one of the three possible escape solutions.

Finally, the third subset contains \emph{injection actions} $\mathcal{A}_{inj}$ specifically designed to attempt the capture of the flag via SQL injection. These actions take the form of the general solution:
\[
\mathcal{A}_{inj} = \{
[Escape] \spc \mathtt{UNION \spc SELECT} \spc [FColumns] \spc \mathtt{FROM \spc Flagtable} \},
\]
where $[FColumns]$ corresponds to the repetition of the string $\mathtt{flag}$ a number of times between $1$ and $N_c$.
The cardinality of this subset is $\left| \mathcal{A}_{inj} \right| = 3 \cdot 5 = 15$, that is five actions for each one of the three possible escape solutions.

The total amount of action $\left| \mathcal{A} \right|$ is given by the union of these three partitions $\left| \mathcal{A}_{esc} \cup \mathcal{A}_{col} \cup \mathcal{A}_{inj} \right| = 51$.
Since the solution of the problem belongs to this set, an agent could just try to solve the SQL injection problem by blind guessing, iterating over all the actions in $\mathcal{A}$. In this case, the expected number of attempts before successfully capturing the flag would be $\frac{\left| \mathcal{A} \right|}{2} = 25.5$.
However, like a human pen-tester, a smart agent would take advantage of the structure in the action set $\mathcal{A}$: several actions overlap, and by first discovering which escape character works in the pre-generated SQL query, it is possible to reduce the space of reasonable actions by two thirds. Thus, a proper balance of exploration and exploitation may lead to a much more effective strategy. An optimal policy would consist of a number of steps proportional to the expected number of actions necessary to find the right escape character, $\frac{3}{2} = 1.5$, plus the expected number of actions necessary to perform an exploitation action with the right number of columns, $\frac{5}{2} = 2.5$; because of the possible overlap between determining the escape character and evaluating the number of columns, we estimate the lower bound on the expected number of steps of an optimal policy to be between $5$ and $4$.

\paragraph*{SQL responses}
Whenever the agent selects an action $a \in \mathcal{A}$, the corresponding SQL statement is sent to the web server, embedded in the pre-generated SQL query, and forwarded to the database. Since we assumed that the processing of the database is transparent, the database response is then provided to the agent.
For the sake of its attack, the agent discriminates between three types of answers: (i) a \emph{negative answer}, that is, an empty answer (due to an invalid SQL statement not matching the escape characters of the pre-generated query); (ii) a \emph{positive answer}, that is, an answer containing data (due to having submitted a valid SQL statement); (iii) the \emph{flag answer}, that is, an answer containing the flag (meaning that the agent managed to exploit the SQL vulnerability).

\paragraph*{Rewards}
We adopt a simple reward policy to train the agent: the attacker collects a $+10$ reward for capturing the flag while receiving a $-1$ reward for any other action not resulting in the flag's capture.
We chose this reward policy heuristically in order to guarantee a positive return to an agent completing an episode in a reasonable number of steps (an optimal policy would take between 4 and 5 steps), and a negative return to any agent with a sub-optimal strategy taking more than 10 steps to reach the solution. The specific values are, however, arbitrary as the goal of the agent is to maximize the reward; as long as it will receive a strong positive signal for reaching the flag and a negative penalty, the agent will learn, in the long run, a policy that will retrieve the flag in the minimum number of actions.


\paragraph*{Generalization}
We would like to remark that, in each episode, our environment is initialized with a new pre-generated query and a new database structure. Therefore, our agent is not meant to learn one specific solution to a single SQL injection vulnerability. Instead, it is supposed to learn a generic strategy that may flexibly adapt to any vulnerability generated by our environment.

\paragraph*{State space}
Our state will be defined by the set of actions performed as well as the response.
This naive state space definition means that we can have up to $3^{|\mathcal{A}|}$ different states. More on the actual implementation is discussed in the simulation sections.

\subsection{Simulation 1}
In our first simulation, we implemented a simple tabular Q-learning agent, and we trained and tested it in our environment.

\paragraph*{Agent}
Our tabular Q-learning agent tracks all the actions performed and whether their outcome was a \emph{negative} or \emph{positive} answer. Notice that we do not need to track in memory the \emph{flag} answer since it marks the end of an episode.
The game's state is then described by the collection of actions and relative responses, forming the history $h$.
An example of history may be $h=\{ {8}, -{16}, {21}\}$, denoting that the agent has taken action $a_8$ and $a_{21}$, that returned a positive answer, and action $a_{16}$ that returned a negative answer. 
For each possible history, $h$, the agent then maintains a probability distribution over the actions, $Q(h,a)$.

Notice, that even with a modest amount of actions, the cardinality of the state space has an unmanageable size of $2^{\left| \mathcal{A} \right|} = 2^{51}$. To workaround this computational issue, we exploit the fact that a large number of these possible histories are not consistent (i.e., we can not have positive and negative answers for actions with the same escape characters) and will never be explored; we then rely on using a sparse Q-table instantiated just-in-time, where entries of the Q-table are initialized and stored in memory only when effectively encountered.

\paragraph*{Setup}
We run our environment using $N_c = 5$ as already described. For statistical reasons, we train $10$ agents using a discount factor $\gamma=0.9$, an exploration rate $\epsilon=0.1$, and a learning rate $\eta=0.1$. We run each agent on $10^6$ episodes.

\begin{figure}
\centering
\begin{subfigure}{.5\textwidth}
  \centering\captionsetup{width=.8\linewidth}
  \includegraphics[width=.6\linewidth]{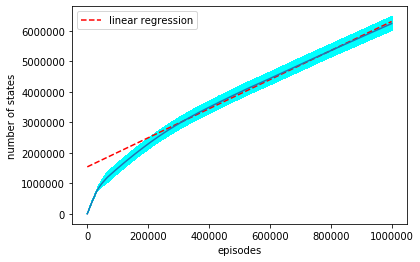}
  \caption{Number of states instantiated by the agent. The dark blue line represents the average computed over the 10 agents, the blue shaded area represents the standard deviation. The red dashed line is a linear regression on the domain $[2\cdot10^5,10^6]$.}
  \label{fig:sim1_nstates}
\end{subfigure}%
\begin{subfigure}{.5\textwidth}
  \centering\captionsetup{width=.8\linewidth}
  \includegraphics[width=.6\linewidth]{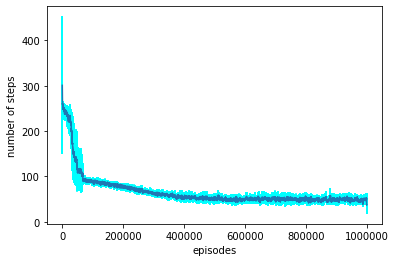}
  \caption{Number of steps per episode. The number of steps for each agent is first smoothed using a 1000-step window; the dark blue line represents the average computed over the 10 agents, the blue shaded area represents the standard deviation.}
  \label{fig:sim1_nsteps}
\end{subfigure}
\caption{Simulation 1 - training.}
\end{figure}

\paragraph*{Results and analysis}
First of all we analyze the dynamics of learning.
Figure \ref{fig:sim1_nstates} shows the number of states instantiated by the agents in their Q-tables during training. We can clearly notice two different learning phases: an exponential growth at the very beginning, while the agents discover new states; and then a longer phase characterized by a linear growth, in which the agents keep discovering new states spurred by their exploration rate parameter. Figure \ref{fig:sim1_nstates} shows that, starting at around episode $2\cdot10^5$, this growth is well captured by linear regression, suggesting a slowing down in learning.
Figure \ref{fig:sim1_nsteps} reports the number of steps per episode; the plot is smoothed by averaging together $1000$ episodes in order to make the trend more apparent. At the very beginning, starting with a uniform policy, the agents act randomly, and this implies that a large number of actions is required to find the solution; at the very beginning, both the average number of actions and the standard deviation is very high, highlighting the purely random behaviour of the agents. Notice that the number of steps is far higher than the cardinality of the action space, because the initial agents may re-sample the same action multiple times; this may seem extremely ineffective and irrational, but notice that the agents have no way to know that the order of actions does not matter. By the end of the training, the number of actions has decreased considerably, although the agents are still taking random actions from time to time due to their exploratory policy; notice that, as soon as a random exploratory action is taken, an agent may find itself in an unknown state in which it has not yet learned how to behave optimally; as such, every time an agent act in an exploratory way multiple new states may be added to its Q-table (as shown by Figure \ref{fig:sim1_nstates}) and multiple steps may be necessary to get to the solution.

\begin{figure}
\centering
\begin{subfigure}{.3\textwidth}
  \centering
  \includegraphics[width=.8\linewidth]{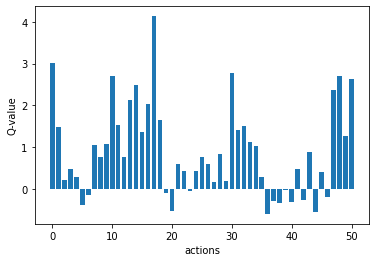}
  \caption{Plot of $Q(\{1\},\cdot)$.}
  \label{fig:sim1_Q_+1}
\end{subfigure}%
\begin{subfigure}{.3\textwidth}
  \centering
  \includegraphics[width=.8\linewidth]{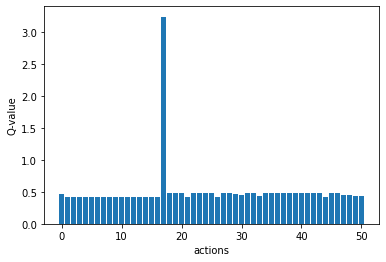}
  \caption{Plot of $Q(\{-1\},\cdot)$.}
  \label{fig:sim1_sim1_Q_-1}
\end{subfigure}
\caption{Simulation 1 - Q-tables.}
\end{figure}

Using a Q-table also allows us to introspect the behaviorual policy of an agent by reading out the entries of the table. For instance, Figure \ref{fig:sim1_Q_+1} and Figure \ref{fig:sim1_sim1_Q_-1} illustrates two entries in the Q-table of an agent, respectively for state $h=\{1\}$ (the agent has taken only action $a_1$ and it has received a positive answer) and $h=\{-1\}$ (the agent has taken only action $a_1$ and it has received a negative answer).
In the case of Figure \ref{fig:sim1_Q_+1}, the policy of the agent has evolved from the original uniform distribution to a more complex multimodal distribution. In particular, a large amount of the probability mass is distributed between action $a_{12}$ and action $a_{17}$, with the most likely option being action $a_{17}$ ($\mathtt{" \spc UNION \spc SELECT \spc flag,flag,flag,flag,flag \spc FROM \spc Flagtable}$); the set of action between action $a_{12}$ and action $a_{17}$ correspond indeed to the set of potentially correct queries, consistent with the escape character discovered by action $a_1$. However, probability mass still remains on necessarily wrong alternatives; further training would likely lead to removal of this probability mass.
In the case of Figure \ref{fig:sim1_sim1_Q_-1}, we observe an almost-deterministic distribution, hinting at the fact that the agent has worked out an optimal option; reasonably, after the failure of action $a_1$, the agent almost certainly will opt for action $a_{18}$ which allows it to test a different escape character.
We hypothesize that such a deterministic behaviour is likely due to a quick build-up of the probability of action $a_{18}$ in early training; as action $a_{18}$ turned out to be a reasonable and rewarding choice, the agent entered in a self-reinforcing loop in which, whenever in state $h=\{-1\}$, it chose action $a_{18}$, and action $a_{18}$ got reinforced after achieving the objective.


\begin{figure}
\centering
\begin{subfigure}{.4\textwidth}
  \centering\captionsetup{width=.8\linewidth}
  \includegraphics[width=.8\linewidth]{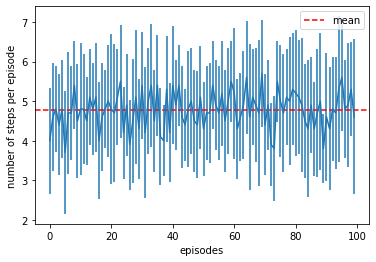}
  \caption{Number of steps per episode. The dark blue line represents the average computed over the 10 agents, the light blue lines represent the standard deviation. The red dashed line is the mean across all the episodes.}
  \label{fig:sim1_test1}
\end{subfigure}%
\begin{subfigure}{.6\textwidth}
  \centering\captionsetup{width=.8\linewidth}
  \includegraphics[width=.8\linewidth]{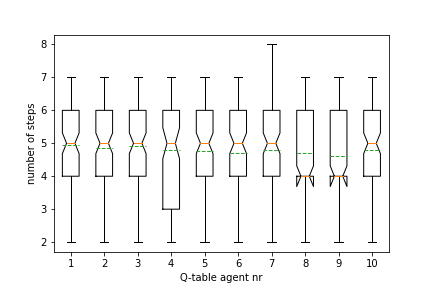}
  \caption{Notch plot of the 10 tabular Q-learning agents performance in number of steps. The orange lines and the green dashed lines represent respectively the median and mean of steps. The notches around the median give a 95\% confidence interval for the median. The box around the median identifies the (25th-75th)-percentile of the distribution, with the top and bottom box giving each the 25\% of the probability mass above and below the median. The whiskers at the top and bottom show the remaining probability mass above and below the median.}
  \label{fig:sim1_1_notch1}
\end{subfigure}%
\caption{Simulation 1 - testing.}
\end{figure}

Finally, we analyzed the behaviour of the agents at test time, by setting their exploration parameter to zero, $\epsilon=0$, and running $100$ further episodes. This emulates the actual deployment of an agent in a real scenario in which we do not want the agent to explore anymore, but just to aim directly for the flag in the most efficient way possible. Notice that, while setting the exploration parameter $\epsilon$ to zero, we still keep the learning parameter $\eta$ different from zero; this, again, is meant to reflect a real scenario, in which an agent keeps learning even at deployment time, ideally to capture possible shifts in the environment in which it operates.
Figure \ref{fig:sim1_test1} shows the number of steps per episode. The blue lines show mean and standard deviation in the number of steps for our 10 agents, while the red dashed line provides a global mean of number of steps per episode across all the agents and all the episodes.
This average number of actions is very close to the theoretical expectation that we identified between $4$ and $5$. Notice that, of course, the expectation holds in a statistical sense: longer episodes taking $6$ or $7$ steps are balanced by fortuitous episodes as short as $2$ steps where the agent guessed by chance the right SQL injection query.
A more detailed overview of the statistical performance of each agent is provided by the notch plot in Figure \ref{fig:sim1_1_notch1}. Each notch provides information on the main statistics about the distribution of the number of steps taken by each of the 10 trained tabular Q-learning agents.
All the agents perform indeed similarly. The median number of steps is slightly better for agents 8 and 9, which get closer to the lower bound, but this small value may be an effect of the small number of tests (100), as we see that its mean is very similar to the others.
The notches around agents 8 and 9 are outside the bottom box; this is because the middle bottom 25\% of the data all require exactly 4 steps, so part of our confidence interval is outside the bottom box.
On the other hand, agent 7 is the only one that may take more than 7 steps to completion. This is a sub-optimal result as a the trivial policy of using 2 exploratory actions and 5 injection guesses would be sufficient to capture the flag; however, this sub-optimal behaviour requiring 8 steps happens only in 2\% of the episodes, leading to hypothesize that they constitute outlier behaviours.

\paragraph*{Discussion}
The results of this simulation show that even a simple reinforcement learning agent based on a tabular Q-learning algorithm can successfully develop an effective strategy to solve our SQL injection problem. Such an agent relies on minimal prior knowledge provided by the designer. We showed, through an analysis of the learning dynamics at training time that a tabular Q-learning agent can discover a meaningful strategy by pure trial and error, and we demonstrated that, at test time, such an agent can reach a performance close to the theoretical optimum.
However, although using a table to store the Q-value function has allowed us to carry a close analysis of the learning dynamics of the agent, it is clear that this approach has poor scalability. We observed how the Q-table keeps growing during all the episodes we ran, and it is immediate to infer that, if the action or state space were to increase, this approach would be infeasible. In the next simulation we sacrifice interpretability in order to work-around the issue of scalability.

\subsection{Simulation 2}
In this simulation, we deploy a more sophisticated agent, a deep Q-learning agent, to tackle the same learning problem as in Simulation 1.

\paragraph*{Environment} We implement the same environment as Simulation 1 as a standard OpenAI environment\footnote{\url{https://github.com/openai/gym}}.

\paragraph*{Agent} We instantiate a deep Q-learning agent using a standard implementation from the \emph{stablebaselines} library\footnote{\url{https://github.com/DLR-RM/stable-baselines3}}.

\paragraph*{Setup} We use the same environment settings as Simulation 1.
As before we train $10$ agents. Because of the definition of the DQN algorithm, we specify the length of training in terms of steps, and not episodes. We then train the agents for $10^6$ steps, corresponding approximately to $10^5$ episodes, in order to guarantee that the DQN agents would not unfairly be trained for longer than the tabular Q-learning agent. We use the default values for the hyperparameters of the DQN agents, although we also consider increasing the batch size from the default $32$ to $51$ in order to increase the likelihood for the agent to observe positive rewards in its batch.



\begin{figure}
\centering
\begin{subfigure}{.4\textwidth}
	\centering\captionsetup{width=.8\linewidth}
  \includegraphics[width=.8\linewidth]{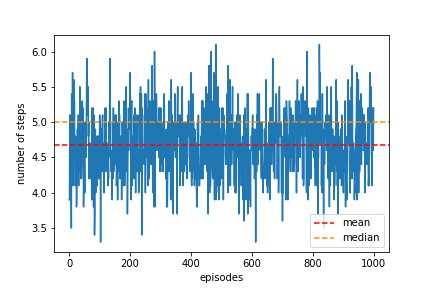}
  \caption{Number of steps per episode. The blue line represents the average computed over the 10 agents. The red dashed line is the mean across all the episodes, the yellow line the median.}
  \label{fig:sim2_b51_test1}
\end{subfigure}%
\begin{subfigure}{.4\textwidth}
  \centering\captionsetup{width=.8\linewidth}
  \includegraphics[width=.8\linewidth]{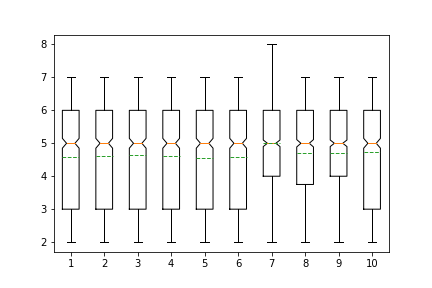}
  \caption{Notch plot of the number of steps for each of the 10 different agents. For the meaning of the plot, refer to Figure \ref{fig:sim1_1_notch1}.}
  \label{fig:sim2_b51_test2}
\end{subfigure}
\caption{Simulation 2 - testing.}
\end{figure}

\paragraph*{Results and analysis}
We start evaluating the performance of the DQN agents at test time, computed on $1000$ test episodes with no exploration, in terms of number of steps necessary to achieve the solution.
Figure \ref{fig:sim2_b51_test1} shows the performance of the DQN agents trained with a batch size of 51. These agents successfully learned competitive policies.
As in the case of the tabular Q-learning agent, we can observe that the overall number of steps averaged over all the episodes and all the agents (red dashed line) settles between 4 and 5 steps, again close to the lower bound identified earlier.
A closer look at the performance of each agent is provided by the notch plot in Figure \ref{fig:sim2_b51_test2}. 
The median and mean are always equal or less than 5 steps. A single DQN agent sometimes achieves a solution in a sub-optimal number of steps (8 steps), as in the case of the tabular Q-learning agent.

When training the DQN agents with a default batch size of 32, some of our vanilla DQN agents were still able to learn a satisfactory strategy, while others failed in their learning task. A detailed analysis of our results is provided in \ref{app:furtherresults}. It is clear that using what turned out to be a sub-optimal batch size of 32 made learning more challenging; agents could still learn but this may require longer training time in order to successfully learn optimal policies. 

\begin{figure}
\centering
\begin{subfigure}{.5\textwidth}
	\centering\captionsetup{width=.8\linewidth}
	\includegraphics[width=.6\linewidth]{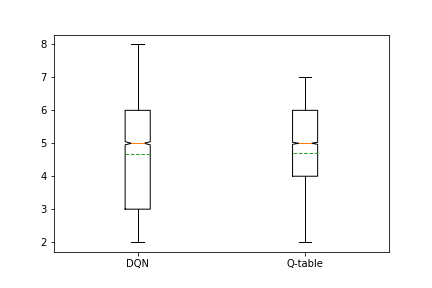}
	\caption{Notch plot of the number of steps taken by the DQN agents compared with the tabular Q-learning agents. For the meaning of the plot, refer to Figure \ref{fig:sim1_1_notch1}.}
	\label{fig:sim2_b51_test4}
\end{subfigure}%
\begin{subfigure}{.5\textwidth}
  \centering
  \includegraphics[width=.8\linewidth]{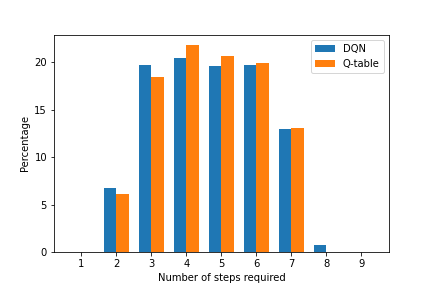}
  \caption{Bar plot comparing the number of steps (in percentage) taken by the DQN agents and the tabular Q-learning agents.}
  \label{fig:sim2_b51_test5}
\end{subfigure}
\caption{Comparison between the DQN and the tabular Q-learning models.}
\end{figure}

To assess the difference between the tabular Q-learning agents and the the DQN agents, we run a last direct comparison by re-training the agents (tabular Q-learning agent and DQN agent with batch size 51) and testing them on 1000 SQL environments.
Figure \ref{fig:sim2_b51_test4} captures this direct comparison between the DQN agent and the tabular Q-learning agents.
The notch plot shows the distributions of the number of steps aggregated over the 10 agents we trained. The two agents perform similarly in terms of mean and median; however, the notch box of the DQN agent has a larger support, suggesting that certain attacks may be completed in fewer steps; while 50\% of the attacks of the tabular Q-learning agent are completed in 4 to 6 steps, 50\% of the attacks of the DQN agent are completed in 3 to 6 steps. Also, note that the retrained tabular Q-learning agents were all quite good, and none of them ever used 8 steps, while some of the DQN agents used 8 steps. This strenghtens the hypothesis that these behaviours may be treated as outliers. 
A more detailed view of the actual distribution of the number of steps taken to reach a solution is provided in Figure \ref{fig:sim2_b51_test5}; the DQN agents present a higher proportions of solutions consisting of only two or three steps, while the tabular Q-learning agents have higher proportions of solution with 4, 5, 6, or 7 steps; as pointed out by the notch plot, only the DQN agents use 8 steps in few instances.

Beyond contrasting the performance of the tabular Q-learning and the DQN agents, an instructive comparison is in terms of the size of the learned models. As expected, the final deep Q-learning model is substantially smaller than the one learned by the tabular Q-learning agent. At the end of the training, the Q-table instantiated by the tabular Q-learning model had a size in the order of a gigabyte (consistent with a table having around $1.75\cdot10^6$ entries of $51$ floats), while the network instantiated by the deep Q-learning model had a constant size in the order of hundreds of kilobytes (consistent with the set of weights of its neural network).

\paragraph*{Discussion}
The deep Q-learning agents were able to learn a good strategy for the SQL injection problem, while, at the same time, provide a solution to the space constraints imposed by the instantiation of an explicit Q-table. Using a deep neural network allows to scale up the problems we may consider, although, on the negative side, relying on black-box neural networks has prevented us from easily examining the inner dynamics of the model as we did for the tabular Q-learning agent. Nonetheless, such an agent may constitute a good starting point for tackling more realistic SQL injection challenges.

\section{Ethical considerations}
The development of an autonomous agent able to perform the probing of a system and the exploitation of a potential SQL injection vulnerability carries inevitable risks of misuse. In this research, the authors were concerned with the development of proof-of-concept agents that may be of use for legitimate penetration testing and system assessment; as such, all the agents were trained to exploit the vulnerability simply by obtaining some special data (flag) inside the database. Although the models learned may not yet cope with real-world scenarios, it is not far-fetched to conceive of future malicious uses for such agents. The authors do not support the use of their research for such aims, and condemn the use of autonomous agents for unethical and illegitimate purposes, especially in a military domain\footnote{\url{https://futureoflife.org/open-letter-autonomous-weapons/}}.

\section{Conclusion}
In this paper, we showed how the problem of exploiting SQL injection vulnerability may be expressed as a reinforcement learning problem.
We considered a simplified SQL injection problem, we formalized it, and we instantiated it as an environment for reinforcement learning.
We then deployed Q-learning agents to solve the problem, showing that both interpretable and straightforward tabular Q-learning agents and more sophisticated deep Q-learning agents can learn meaningful strategies.
These results provide proof-of-concept support to the hypothesis that reinforcement learning agents may be used in the future to perform penetration testing and security assessment.
However, our results are still preliminary; real-world cases of SQL injection present higher levels of complexity, which constitute a significant challenge for both modeling and learning.

Future work may be directed at considering more realistic setups, as well as deploying more sophisticated agents.
The current solution can be improved by considering a larger (possibly combinatorial) action space, or by extending the types of vulnerabilities (e.g., error-based or time-based SQL injection). Alternatively, a more realistic model may be produced by providing the agent with non-preprocessed answers from the web server in the form of HTML pages. All these directions represent important development that would allow us to model more realistic settings and train more effective autonomous agents.


\bibliographystyle{elsarticle-num}
\bibliography{ctf}

\begin{thebibliography}{10}
\expandafter\ifx\csname url\endcsname\relax
  \def\url#1{\texttt{#1}}\fi
\expandafter\ifx\csname urlprefix\endcsname\relax\def\urlprefix{URL }\fi
\expandafter\ifx\csname href\endcsname\relax
  \def\href#1#2{#2} \def\path#1{#1}\fi

\bibitem{mnih2015human}
V.~Mnih, K.~Kavukcuoglu, D.~Silver, A.~A. Rusu, J.~Veness, M.~G. Bellemare,
  A.~Graves, M.~Riedmiller, A.~K. Fidjeland, G.~Ostrovski, et~al., Human-level
  control through deep reinforcement learning, Nature 518~(7540) (2015)
  529--533.

\bibitem{silver2017mastering}
D.~Silver, J.~Schrittwieser, K.~Simonyan, I.~Antonoglou, A.~Huang, A.~Guez,
  T.~Hubert, L.~Baker, M.~Lai, A.~Bolton, et~al., Mastering the game of {G}o
  without human knowledge, Nature 550~(7676) (2017) 354.

\bibitem{vinyals2019grandmaster}
O.~Vinyals, I.~Babuschkin, W.~M. Czarnecki, M.~Mathieu, A.~Dudzik, J.~Chung,
  D.~H. Choi, R.~Powell, T.~Ewalds, P.~Georgiev, et~al., Grandmaster level in
  {S}tarcraft {II} using multi-agent reinforcement learning, Nature 575~(7782)
  (2019) 350--354.

\bibitem{erdodi2020agent}
L.~Erdodi, F.~M. Zennaro, The agent web model--modelling web hacking for
  reinforcement learning, arXiv preprint arXiv:2009.11274.

\bibitem{sarraute2013penetration}
C.~Sarraute, O.~Buffet, J.~Hoffmann, Penetration testing== {POMDP} solving?,
  arXiv preprint arXiv:1306.4714.

\bibitem{bland2020machine}
J.~A. Bland, M.~D. Petty, T.~S. Whitaker, K.~P. Maxwell, W.~A. Cantrell,
  Machine learning cyberattack and defense strategies, Computers \& security 92
  (2020) 101738.

\bibitem{ghanem2020reinforcement}
M.~C. Ghanem, T.~M. Chen, Reinforcement learning for efficient network
  penetration testing, Information 11~(1) (2020) 6.

\bibitem{pozdniakov2020smart}
K.~{Pozdniakov}, E.~{Alonso}, V.~{Stankovic}, K.~{Tam}, K.~{Jones}, Smart
  security audit: Reinforcement learning with a deep neural network
  approximator, in: 2020 International Conference on Cyber Situational
  Awareness, Data Analytics and Assessment ({CyberSA}), 2020, pp. 1--8.

\bibitem{zennaro2020modeling}
F.~M. Zennaro, L.~Erdodi, Modeling penetration testing with reinforcement
  learning using capture-the-flag challenges and tabular {Q}-learning, arXiv
  preprint arXiv:2005.12632.

\bibitem{CyberGrand}
D.~Fraze, Cyber grand challenge ({CGC}),
  \url{https://www.darpa.mil/program/cyber-grand-challenge}, accessed:
  2020-05-09 (2016).

\bibitem{sutton2018reinforcement}
R.~S. Sutton, A.~G. Barto, Reinforcement learning: An introduction, MIT press,
  2018.

\bibitem{StasinopoulosAutomatic}
A.~Stasinopoulos, C.~Ntantogian, C.~Xenakis, Commix: automating evaluation and
  exploitation of command injection vulnerabilities in web applications,
  International Journal of Information Security.

\bibitem{krizhevsky2012imagenet}
A.~Krizhevsky, I.~Sutskever, G.~E. Hinton, Imagenet classification with deep
  convolutional neural networks, in: Advances in neural information processing
  systems, 2012, pp. 1097--1105.

\bibitem{vaswani2017attention}
A.~Vaswani, N.~Shazeer, N.~Parmar, J.~Uszkoreit, L.~Jones, A.~N. Gomez,
  {\L}.~Kaiser, I.~Polosukhin, Attention is all you need, in: Advances in
  neural information processing systems, 2017, pp. 5998--6008.

\bibitem{russell2018automated}
R.~Russell, L.~Kim, L.~Hamilton, T.~Lazovich, J.~Harer, O.~Ozdemir,
  P.~Ellingwood, M.~McConley, Automated vulnerability detection in source code
  using deep representation learning, in: 2018 17th IEEE International
  Conference on Machine Learning and Applications (ICMLA), IEEE, 2018, pp.
  757--762.

\bibitem{tobiyama2016malware}
S.~Tobiyama, Y.~Yamaguchi, H.~Shimada, T.~Ikuse, T.~Yagi, Malware detection
  with deep neural network using process behavior, in: 2016 IEEE 40th annual
  computer software and applications conference (COMPSAC), Vol.~2, IEEE, 2016,
  pp. 577--582.

\bibitem{skaruz2007recurrent}
J.~Skaruz, F.~Seredynski, Recurrent neural networks towards detection of sql
  attacks, in: 2007 IEEE International Parallel and Distributed Processing
  Symposium, IEEE, 2007, pp. 1--8.

\bibitem{joshi2014sql}
A.~Joshi, V.~Geetha, Sql injection detection using machine learning, in: 2014
  International Conference on Control, Instrumentation, Communication and
  Computational Technologies (ICCICCT), IEEE, 2014, pp. 1111--1115.

\bibitem{singh2015Sql}
G.~Singh, D.~Kant, U.~Gangwar, A.~P. Singh, Sql injection detection and
  correction using machine learning techniques, in: Emerging ICT for Bridging
  the Future - Proceedings of the 49th Annual Convention of the Computer
  Society of India (CSI), Vol.~1, Springer, 2015, pp. 435--442.

\bibitem{ogbomon2017applied}
S.~O. Uwagbole, W.~J. Buchanan, L.~Fan, Applied machine learning predictive
  analytics to sql injection attack detection and prevention, in: 2017
  IFIP/IEEE Symposium on Integrated Network and Service Management (IM), IEEE,
  2017, pp. 1087--1090.

\bibitem{ross2018sql}
K.~Ross, Sql injection detection using machine learning techniques and multiple
  data sources,
  \url{https://scholarworks.sjsu.edu/cgi/viewcontent.cgi?article=1649&context=etd_projects},
  accessed: 2021-02-15 (2018).

\bibitem{hasan2019detection}
M.~Hasan, Z.~Balbahaith, M.~Tarique, Detection of sql injection attacks: A
  machine learning approach, in: 2019 International Conference on Electrical
  and Computing Technologies and Applications (ICECTA), Ras Al Khaimah, United
  Arab Emirates, IEEE, 2019, pp. 1--6.

\bibitem{tang2020detection}
P.~Tang, W.~Qiu, Z.~Huang, H.~Lian, G.~Liu, Detection of sql injection based on
  artificial neural network, Knowledge-Based Systems 190.

\end{thebibliography}

\appendix
\setcounter{figure}{0}
\section{Additional Results \label{app:furtherresults}}

\subsection*{Simulation 2}

\begin{figure}
\centering
\begin{subfigure}{.4\textwidth}
	\centering\captionsetup{width=.8\linewidth}
  \includegraphics[width=.8\linewidth]{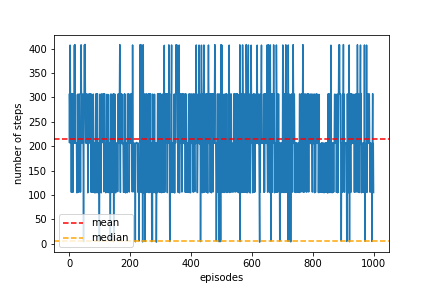}
  \caption{Number of steps per episode. The blue line represents the average computed over the 10 agents. The red dashed line is the mean across all the episodes, the yellow line the median.}
  \label{fig:sim2_b32_test1}
\end{subfigure}
\begin{subfigure}{.4\textwidth}
  \centering\captionsetup{width=.8\linewidth}
  \includegraphics[width=.8\linewidth]{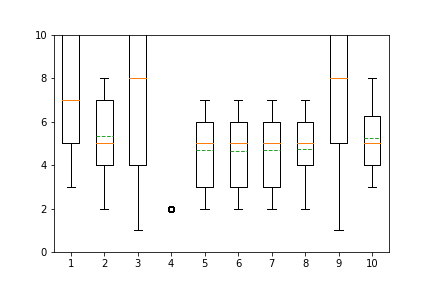}
  \caption{Notch plot of the number of steps for each of the 10 different agents. The $y$-axis has been clipped at $10$. For the meaning of the plot, refer to Figure \ref{fig:sim1_1_notch1}.}
  \label{fig:sim2_b32_notch_test6}
\end{subfigure}\\%
\begin{subfigure}{.4\textwidth}
  \centering
  \includegraphics[width=.8\linewidth]{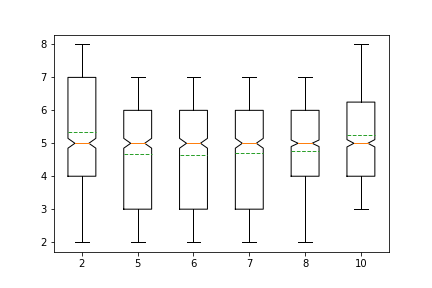}
  \caption{Notch plot of the number of steps for each of the 6 well-behaving agents. For the meaning of the plot, refer to Figure \ref{fig:sim1_1_notch1}.}
  \label{fig:sim2_b32_notch_test4}
\end{subfigure}
\caption{Simulation 2 - testing on DQN agents trained using a batch size of 32.}
\end{figure}

In this section we report full results for the DQN agents trained with batch size 32.
Following the standard protocol, we trained 10 agents and tested their performance on a 1000 SQL environments.
Figure \ref{fig:sim2_b32_test1} shows the mean and median number of steps computed over the 10 agents and 1000 episodes. While the low median of 6 proves that the majority of episodes is solved in a limited number of steps, the very high mean of 214.4 highlights that, even at the end of training, there are still scenarios in which the agents take a large number of steps, likely reaching the step limit of the task; this is probably due to the agent finding itself in unforeseen states and ending up in a loop. It is clear that, in this case, while the agent has learned something about the environment (witnessed by the low median), the training has been insufficient to learn a complete and reliable policy.
Figure \ref{fig:sim2_b32_notch_test6} provide a better insight in this failure, showing a notch plot of the number of steps for each of the 10 different agents. We immediately see that the unsatisfactory results observed in Figure \ref{fig:sim2_b32_test1} are due to the failure in learning of four agents: the notch boxes of agents 1, 3 and 9 stretches far beyond the limit of the y-axis, indicating that a large number of episodes take more than 10 steps; even worse, for agent 4, we only observe an outlier, while the entire notch lies above the limit of the y-axis. These four agents have not been able to learn good policies at training time, and their bad performance affects the overall mean we computed in Figure \ref{fig:sim2_b32_test1}.
If we simply plot the notch graph for the agents that learned successfully (see Figure \ref{fig:sim2_b32_notch_test4}) we notice that the performances of the six agents that trained succesfully closely resemble the performance of the DQN agents trained with batch size 51 (see Figure \ref{fig:sim1_1_notch1}). Indeed the mean number of steps of this smaller group is now $4.895$, which is quite good, although lower than our tabular Q-learning agent, which used an average of $4.795$ steps.
Overall, this analysis showed that training a DQN agent with a batch size of 32 is still doable, although more challenging; with the same computational budget for training as for the agent trained with batch size 51, there is a higher likelihood that the learned policy will not be optimal, and in certain scenarios will fail.


\end{document}